\documentclass{svjour2}                    
\smartqed  
\usepackage{graphicx}
\usepackage{hyperref}
\usepackage{mathptmx}      
\journalname{Foundations of Physics}
\begin{document}

\title{Measurement-based quantum foundations
}

\titlerunning{Measurement-based quantum foundations}        

\author{Jochen Rau}

\authorrunning{Jochen Rau} 

\institute{J. Rau \at
              Institut f\"ur Theoretische Physik,
Johann Wolfgang Goethe-Universit\"at,
Max-von-Laue-Str. 1, 
60438 Frankfurt am Main,
Germany \\
              \email{jochen.rau@q-info.org}           
}

\date{Received: date / Accepted: date}

\maketitle

\begin{abstract}
I show that quantum theory is the only probabilistic framework that permits arbitrary processes to be emulated by sequences of local measurements.
This supports the view that, contrary to conventional wisdom, measurement should not be regarded as a complex phenomenon in need of a dynamical explanation but rather as a primitive -- and perhaps the only primitive -- operation of the theory.
\end{abstract}

\section{Introduction}
\label{intro}

Of the various axioms of standard quantum mechanics, the measurement postulate marks the most radical departure from classicality.
It entails changes of the quantum state that are discontinuous, in stark contrast to the continuous evolution of ordinary dynamics;
and it raises deep questions -- epitomized by the famous ``Schr\"odinger's cat'' \cite{schrodinger:cat} and ``Wigner's friend'' \cite{wigner:friend} paradoxes -- as to the precise delineation of the border between the quantum and classical realms, between system and observer, and between microscopic and macroscopic.
Its far-reaching conceptual implications are captured most beautifully in a metaphor by John Wheeler \cite{wheeler:lawwithoutlaw}:

\begin{quote}
About the game of twenty questions.
You recall how it goes --- 
one of the after-dinner party sent out of the living room, the others agreeing on a word, the one fated to be questioner returning and starting his questions.
``Is it a living object?'' 
``No.''
``Is it here on earth?''
``Yes.''
So the questions go from respondent to respondent around the room until at length the word emerges:
victory if in twenty tries or less;
otherwise, defeat.

Then comes the moment when we are fourth to be sent from the room.
We are locked out unbelievably long.
On finally being readmitted, we find a smile on everyone's face, sign of a joke or a plot.
We innocently start our questions.
At first the answers come quickly.
Then each question begins to take longer in the answering ---
strange, when the answer itself is only a simple ``yes'' or ``no''.
At length, feeling hot on the trail, we ask,
``Is the word `cloud'?''
``Yes'', comes the reply, and everyone bursts out laughing.
When we were out of the room, they explain, they had agreed not to agree in advance on any word at all.
Each one around the circle could respond ``yes'' or ``no'' as he pleased to whatever question we put to him.
But however he replied he had to have a word in mind compatible with his own reply ---
and with all the replies that went before.
No wonder some of those decisions between ``yes'' and ``no'' proved so hard!
\end{quote}

In Wheeler's game the word does not already exist ``out there'' --- 
rather, information about the word is only brought into being through the questions raised.
While the answers given are internally consistent, and eventually converge to some word, the outcome is influenced by the questioner:
A different sequence of questions will generally lead to a different word.
This constitutes a peculiar form of agent-dependency in that the conclusions drawn depend on the questions asked.
It is distinct from another form of agent-dependency, known from Bayesian probability theory, where conclusions drawn from experimental data generally depend on the agent's prior expectation \cite{fuchs:priors}.
While Wheeler's metaphor should obviously be taken with a grain of salt, it captures one key aspect of quantum theory:
There is no longer a preexisting reality that is merely revealed, rather than influenced, by the act of measurement.
The image of reality that emerges through acts of measurement reflects as much the history of intervention as it reflects the external world \cite{mermin:bell}.

Despite -- or perhaps because of -- the measurement postulate's pivotal role in demarcating the quantum from the classical view of the world many researchers have shown a strong desire, and deemed it possible, to either dispense with the measurement process entirely (as in the many-worlds interpretation \cite{everett:manyworlds}) or at least to reduce it to a less intimidating dynamical process, by means of microscopic collapse models \cite{grw:collapse} and possibly invoking the aid of some stochastic decoherence-inducing environment \cite{zurek:decoherence}. 
Yet in my view such attempts are ultimately misguided.
They merely shift the conceptual challenge to a different level;
and they avoid a square confrontation with what is perhaps the core principle of quantum theory, and with its profound philosophical ramifications.

More recently, alternative views have emerged which suggest that measurement is indeed a primitive process independent of, and possibly even more fundamental than, unitary evolution.
Quantum Bayesianism \cite{caves:quantumasbayes,timpson:bayesianism,fuchs:route},
the quantum version of the homonymous paradigm in classical probability theory \cite{bernardo:book,jaynes:book},
views probabilities -- and hence the quantum state -- as embodying some agent's knowledge about, rather than an objective property of, a physical system \cite{hartle:individual};
so the discontinuous state change brought about by measurement, other than being some mysterious ``collapse'', is but a Bayesian updating that reflects the agent's learning \cite{bub:degrees}.
Within the field of quantum computer science, too, the concept of measurement is unproblematic and in fact forms the foundation of the entire subject \cite{mermin:praise}.
Moreover, while it has proven difficult -- if not outright impossible -- to model the measurement process in terms of unitary evolution (involving modelling of the measurement apparatus and possibly some stochastic environment) the converse is in fact true:
To within any given finite accuracy, every unitary map (in fact every completely positive map) can be emulated by a sequence of measurements \cite{rudolph:emulate}.
This is illustrated by the power of measurement-based quantum computation,
where arbitrary quantum algorithms can be implemented by sequences of measurements only;
such measurement-based quantum computation is just as universal as the network model based on unitary gates \cite{raussendorf:oneway,briegel:measbased}.
In short, while it is not possible to reduce measurement to unitaries, unitaries can certainly be reduced to measurement.
Relational formulations of quantum theory \cite{page:evolution,wootters:time,rovelli:time,poulin:toy} and evidence for the possible ``timelessness'' of quantum gravity \cite{barbour:timeless,hellmann:multiple} even suggest that it may in fact be ``time'' and hence dynamics, rather than measurement, that should be dispensed with at the most fundamental level.

In this paper I address the question whether the central role of the quantum measurement process, rather than being a conceptual nuisance in need of explanation, may actually serve as the logical starting point, i.e., constitute \textit{the} foundational principle for quantum theory.
In order to answer the question in the affirmative I must show that some small set of assumptions about the measurement process plus some basic consistency of inferential reasoning will suffice to recover the full mathematical apparatus -- complex Hilbert space -- of quantum theory.
Such a reconstruction of quantum theory \cite{grinbaum:reconstruction} will be one further proposal in a long history of different approaches (for a recent overview, see Ref. \cite{rau:qvc}).
Ideally, it will be of a simplicity and clarity similar to the derivation of the mathematical apparatus of special relativity -- Lorentz transformations -- from just the physical principle that the speed of light be constant in all reference frames \cite{zeilinger:foundational}.

In Section \ref{postulates} I will propose the central measurement postulate;
in Section \ref{derivation} I will outline the key arguments leading from this postulate to complex Hilbert space;
and in Section \ref{summary} I will conclude with some final remarks.

\section{Central postulate}
\label{postulates}

Before laying down the central measurement postulate I wish to emphasize that I view quantum theory first and foremost as a probability theory.
Like classical probability theory it has as its subject propositions (represented by subspaces of Hilbert space) and their possible logical relations such as implication (embedding) or mutual exclusion (orthogonality); 
and its principal \textit{raison d'\^etre} is to provide a set of rules for assigning to these propositions probabilities in a way which ensures consistency of inferential reasoning. 
The dynamics of a quantum system as described by, say, the Schr\"odinger equation is a subsequent add-on and typically involves the representation of some specific spacetime symmetry group --- in the case of the Schr\"odinger equation, the Galileo group.
Here, however, I shall not regard dynamics but focus exclusively on the probabilistic core.

Underlying quantum theory is a proposition system whose structure is weaker than that of classical logic in one important respect:
In contrast to the latter not all propositions are jointly decidable;
so the Boolean operations ``and'', ``or'' are no longer defined for arbitrary pairs of propositions but for jointly decidable pairs only.
As a consequence, the mathematical structure of the proposition system is no longer that of a distributive lattice \cite{birkhoff:logic,jauch:book} but, more weakly, an orthomodular poset or (equivalently) orthoalgebra \cite{wilce:orthoalgebras}.
It has a well-defined dimension given by the maximum number of mutually exclusive propositions. 
I will henceforth restrict my attention to systems whose dimension is finite.

As for probabilites, these add up whenever two mutually exclusive (and hence also jointly decidable) propositions are concatenated with a Boolean ``or'' operation.
The state of a system is tantamount to an exhaustive list of probabilities for all propositions.
Arbitrary convex combination of such probability tables yields another allowed probability table, as does rescaling with any factor less than one
(states need not be normalised);
so states form a convex cone.
States may be mixed (whenever they can be written as a convex combination of other states), or else pure.
Some propositions (corresponding in quantum theory to one-dimensional subspaces of Hilbert space) are most accurate in the sense that they are not implied by any other proposition.
There is a one-to-one map from these most accurate propositions to pure states:
Such propositions, if true, encompass the entire belief structure and hence must correspond to a unique state;
this state is necessarily pure. 

When updating probabilities after a measurement, consistency of inferential reasoning dictates that different ways of using the same information lead to the same conclusions irrespective of the particular path chosen.
Like in classical Bayes\-ian probability theory this consistency requirement implies a Bayes rule \cite{cox:probability,jaynes:book}, the latter now being restricted to pairs of propositions that are jointly decidable. 
Given maximal prior information, i.e., the truth of some most accurate proposition about a system, learning subsequently that some further proposition is true generally entails -- in contrast to the classical case -- a non-trivial update of knowledge.
Unless the two propositions are jointly decidable,
the original most accurate proposition will no longer be true.
Yet the updated knowledge continues to be maximal,
so there will be some other most accurate proposition which is true.

Upon composing systems, their dimensions multiply.
Concatenating most accurate propositions that pertain to different constituents (and are hence jointly decidable) with the Boolean ``and'' operation yields a most accurate proposition about the whole.
Yet conversely, not all most accurate propositions about the whole need to arise from such a concatenation;
there may be entanglement.

Following these preliminaries the central postulate can now be formulated as follows:
\textit{Local measurement is the sole fundamental operation.}
In other words, it is always possible to emulate arbitrary processes by sequences of local measurements only.
This breaks down into two distinct requirements:
\begin{itemize}
\item[(i)] 
Every global process on a composite system results from concerted action of local processes on its constituents.
Concerted action, in this context, means that local processes may be correlated, i.e., influenced by information extracted elsewhere and exchanged through classical or quantum channels;
yet beyond such correlations, there are no genuinely ``holistic'' degrees of freedom of a global process.
This possibility to reduce global to local processes, lately termed ``operation locality'' \cite{hardy:foliable}, is the counterpart for processes of the possibility of local tomography for states.
The latter features as an axiom (under various names) in several recent reconstructions of quantum theory \cite{hardy:probtheories,barrett:infoprocessing,dariano:whatisspecial,dakic:quantum}.
It is in fact indispensable for the ability to discover physical laws and make testable predictions:
Only if a theory is locally tomographic do observational data obtained from a sufficiently large sample allow one to assign a unique state to an exchangeable sequence, and hence to infer the properties of a larger system from those of a sample \cite{caves:definettistates,renner:symmetry}.
\item[(ii)]
Locally, for any single constituent of finite size and to within any given finite accuracy,
every process can be emulated by a finite sequence of measurements.
The setups of these measurements need not be deterministic but may be subject to a probability distribution, 
and the latter may in turn be conditioned on the feed forward of preceding measurement outcomes (as is the case in measurement-based quantum computation \cite{raussendorf:oneway,briegel:measbased}). 
\end{itemize}
I shall argue that these two requirements uniquely single out quantum theory in complex Hilbert space.

\section{Derivation of complex Hilbert space}
\label{derivation}

The first part of the central postulate immediately implies that if the theory is described in a Hilbert space, it must be a complex Hilbert space.
Let $S(d)$ denote the minimum number of propositions whose probabilities suffice to specify an arbitrary, not necessarily normalised mixed state of a $d$-dimensional system.
A general process -- corresponding in the quantum case to a completely positive map \cite{kraus:book} -- maps such a fiducial set of $S(d)$ probabilities to $S(d)$ different probabilities, thereby preserving consistency of inferential reasoning (in particular positivity) but not necessarily normalisation;
it is by necessity linear \cite{barrett:infoprocessing}.
Thus the process itself is characterised by $[S(d)]^2$ real parameters.
In order to meet the requirement of operation locality, this number of parameters must satisfy the multiplication law
\begin{equation}
	S(d_A d_B) = S(d_A)\cdot S(d_B)
	\quad.
\end{equation}
Of all Hilbert spaces over a skew field (real, complex or quaternionic) only $d$-dimensional Hilbert space over the complex numbers with $S(d)=d^2$ satisfies this constraint.
In the following, therefore, it will suffice to show that the theory must be described in \textit{some} Hilbert space.

The second part of the central postulate unfolds its power already when applied to one specific process, namely that of steering a system -- without loss -- from a pure state in which some most accurate proposition $e_0$ is true, to another pure state in which a different most accurate proposition $e$ is true.
Such loss-free steering by means of measurements only is the mirror image of the Zeno effect \cite{misra:zeno}.
Yet in order for the theory to exhibit the Zeno effect, it must be smooth in the following double sense:
\begin{itemize}
\item[(i)]
For any system of finite dimension $d$ ($d\ge 2$)
the associated set $X(d)$ of most accurate propositions must be a continuous manifold of non-zero dimension $\dim X(d)>0$.
This manifold is endowed with a natural metric \cite{belinfante:spaces,fivel:interference}
\begin{equation}
	\mbox{dist}(e,f):=\sup_\rho |\mbox{prob}(e|\rho)-\mbox{prob}(f|\rho)|
	\quad,
\end{equation}
compact and connected;
and in the special case of a generalised bit ($d=2$) it is isomorphic to the boundary of the convex set of normalised states\footnote{
or a convex subset thereof, in the hypothetical case (not realised in quantum theory) that the map from most accurate propositions to pure states is not surjective
}
and thus, moreover, simply connected.
(In quantum theory $X(2)$ is isomorphic to the surface of the Bloch sphere.)
This kind of smoothness is assumed, in one form or another, also in other reconstructions of quantum theory \cite{hardy:probtheories,dakic:quantum,wilce:fourandhalf}.
\item[(ii)]
Probabilities that are initially equal to one must not suddenly jump to a lower value upon an infinitesimal displacement.
In mathematical terms, given the initial $e_0\in X(d)$ and any proposition $x$ which in the associated pure state is true with certainty, $\mbox{prob}(x|e_0)=1$, the probability of the latter must change in a continuous fashion:
\begin{equation}
	\forall\,\epsilon>0\ \quad \exists\,\delta>0:
	\ \mbox{prob}(x|e)>1-\epsilon \quad \forall\  e\in {\cal B}(e_0;\delta)
	\quad.
\label{continuity}
\end{equation}
Here ${\cal B}(e_0;\delta)$ denotes an open ball in $X(d)$ of radius $\delta$ (in the metric defined above) around $e_0$.
\end{itemize}

The latter continuity condition implies that the manifold dimension of $X(d)$ grows linearly with $d$,
\begin{equation}
	\dim X(d) = \dim X(2)\cdot (d-1)
	\quad.
\label{xdimension}
\end{equation}
A detailed proof of this dimensional constraint can be found in Ref. \cite{rau:qvc};
here I sketch the main idea for the case $d=3$.
Thanks to continuity, and taking the proposition $x$ in the continuity condition to be two-dimensional, there exists a unique pair $(e',y)$ of mutually exclusive propositions where
(i)
$e'$ is most accurate and implies $x$,
whereas
(ii)
$y$ is two-dimensional and implied by $e$.
The two most accurate propositions $e', e$ can thus be regarded as points either on the same manifold $X(3)$ or on two distinct manifolds $X(2)$ that are associated with reduced two-dimensional theories in which $x$ or $y$ are given as true, respectively.
Specifying $e$ on $X(3)$ is tantamount to specifying first $e'$ on $X(2)$, and hence $y$, given $x$ and then $e$ on $X(2)$ given $y$;
so $\dim X(3)=2\cdot \dim X(2)$.
Iteration of this argument leads to the formula for arbitrary $d$.

In a next step one can employ techniques from group theory to demonstrate that already the above dimensional constraint forces quantum theory to be described in a Hilbert space.
The proposition system, endowed with the mathematical structure of an orthomodular poset or orthoalgebra, gives rise to a group ${\cal G}(d)$ of automorphisms preserving implication and mutual exclusion.
Without loss of generality one may assume that this group acts transitively on the set of most accurate propositions ---
if not, the latter can be decomposed into irreducible components, each of which is the orbit of some most accurate proposition under action of the group.
The manifold $X(d)$ (or any irreducible component thereof) is therefore a homogeneous space \cite{barut:book}
\begin{equation}
	X(d) \sim {\cal G}(d)/[{\cal G}(d-1)\otimes {\cal G}(1)]
	\quad,
\label{isomorphism}
\end{equation}
with the stability group preserving both the given most accurate proposition (of dimension one) and its orthocomplement of dimension $(d-1)$.
Thanks to the above isomorphism all constraints on $X(d)$ readily translate into constraints on ${\cal G}(d)$;
and one can subsequently use well known classification theorems to distill those groups that are permitted.

To begin with, as $X(d)$ is continuous, ${\cal G}(d)$ must be a Lie group;
and $X(d)$ being irreducible, compact and (at least for $d=2$) simply connected, ${\cal G}(d)$ must itself be compact and simple up to factors $U(1)$.
Its Lie group dimension must satisfy
\begin{equation}
	\dim X(d) = \dim {\cal G}(d) - \dim {\cal G}(d-1) - \dim {\cal G}(1)
\end{equation}
which, together with Eq. (\ref{xdimension}), implies the quadratic form
\begin{equation}
	\dim {\cal G}(d) = \dim X(2)\cdot d(d-1)/2 + \dim {\cal G}(1)\cdot d
	\quad.
\end{equation}
Consultation of the classification of compact simple Lie groups \cite{barut:book} reveals that this dimensional constraint restricts ${\cal G}(d)$ to be one of $SO(n d)$, $U(n d)$ or $Sp(n d)$, with $n$ being a positive integer.
All these possibilities correspond to some Hilbert space structure over the reals, complex numbers or quaternions, respectively.
In combination with the first part of the central postulate, this uniquely singles out quantum theory in complex Hilbert space.
\smartqed
\qed

\section{Summary}
\label{summary}

Quantum theory is one of many conceivable alternatives to classical probability theory as a consistent framework for inferential reasoning.
I have argued here that quantum theory is distinguished from all other alternatives by the primacy of measurement:
It is the only probabilistic theory where sequences of local measurements can emulate arbitrary processes.
Conceptually this result suggests that, at least in the finite-dimensional case considered here, quantum theory is the only probabilistic theory that can maintain the appearance of time (permitting arbitrary processes) even when there is no time (dispensing with unitaries at the fundamental level);
and which thus in principle allows for the abolition of time.
Developing this idea further, it appears that a fundamental theory which is strictly relational (and hence without external time) must necessarily be quantum.
As the latter in turn forces one to abandon the notion of a preexisting reality \cite{mermin:bell} 
one may be led to conclude quite generally that relationalism and realism exclude each other.

It is my hope that the findings of the present paper will be of interest in a number of different areas.
They may provide a basis for novel reconstructions of quantum theory;
they may further underpin efforts to demarcate quantum information processing from information processing in other generalized probabilistic theories;
somewhat more speculatively, they might inform future attempts at constructing a strictly relational -- and possibly measurement-based -- theory of quantum gravity;
and finally, they may add a new perspective to the ongoing scientific and philosophical debate about the nature of time and change
\cite{butterfield:time,silagadze:zeno,clockandquantum}.

\begin{acknowledgements}
My thoughts about the subject have benefited from the stimulating environment at both the QTRF5 conference in V\"axj\"o (Sweden) and the Reconstructing Quantum Theory conference at Perimeter Institute.
I thank Marcus Appleby, Giacomo Mauro D'Ariano, Chris Fuchs, Philip Goyal, Alexei Grinbaum, Lucien Hardy, Terry Rudolph, Robert Spekkens and Alex Wilce for helpful discussions,
as well as David Mermin and Cozmin Ududec for pertinent questions about an earlier version of this paper.
\end{acknowledgements}

\bibliographystyle{spphys}       


\end{document}